\documentclass[submission,copyright,creativecommons]{eptcs}
\providecommand{\event}{LFMTP 2021} 
\usepackage{breakurl}             
\usepackage{underscore}           
\usepackage{utf}
\usepackage{xspace}
\usepackage[color]{coqdoc}
\usepackage{amssymb}
\usepackage{graphicx}

\def\Coq{\textsc{Coq}\xspace}
\def\Program{\textsc{Program}\xspace}
\def\Agda{\textsc{Agda}\xspace}
\def\MetaCoq{\textsc{MetaCoq}\xspace}
\def\OCaml{\textsc{OCaml}\xspace}
\def\TemplateCoq{\textsc{Template-Coq}\xspace}
\def\Prop{\coqdockw{Prop}\xspace}
\def\Type#1{\coqdockw{Type}\texttt{@\{}#1\texttt{\}}}
\def\Typea{\coqdockw{Type}\xspace}
\def\PCUIC{\textsc{Pcuic}\xspace}
\def\CertiCoq{\textsc{CertiCoq}\xspace}

\title{Touring the MetaCoq Project \small{(Invited Paper)}}
\author{Matthieu Sozeau
\institute{Inria \& LS2N, Université de Nantes\\France
\email{matthieu.sozeau@inria.fr}
}}
\def\titlerunning{Touring the MetaCoq project}
\def\authorrunning{M. Sozeau}
\begin{document}
\maketitle

\begin{abstract} 
  Proof assistants are getting more widespread use in research and industry to provide certified
  and independently checkable guarantees about theories, designs, systems and implementations. 
  However, proof assistant implementations themselves are seldom verified, although they take a 
  major share of the trusted code base in any such certification effort. In this area, proof 
  assistants based on Higher-Order Logic enjoy stronger guarantees, as self-certified implementations
  have been available for some years. One cause of this difference is the inherent complexity of 
  dependent type theories together with their extensions with inductive types, universe polymorphism and complex sort systems,
  and the gap between theory on paper and practical implementations in efficient programming languages. 
  MetaCoq is a collaborative project that aims to tackle these difficulties to provide the first 
  fully-certified realistic implementation of a type checker for the full calculus underlying 
  the Coq proof assistant. To achieve this, we refined the sometimes blurry, if not incorrect, 
  specification and implementation of the system. We show how theoretical tools from this 
  community such as bidirectional type-checking, Tait-Martin-Löf/Takahashi's confluence 
  proof technique and monadic and dependently-typed programming can help construct the 
  following artifacts:

  \begin{itemize}
  \item a specification of Coq's syntax and type theory, the Polymorphic Cumulative Calculus
   of (Co)-Inductive Constructions (\PCUIC);
  \item a monad for the manipulation of raw syntax and interaction with the Coq system;
  \item a verification of \PCUIC's metatheory, whose main results are the confluence of reduction,
   type preservation and principality of typing;
  \item a realistic, correct and complete type-checker for \PCUIC;
  \item a sound type and proof erasure procedure from \PCUIC to untyped λ-calculus, i.e.,
   the core of the extraction mechanism of Coq.
  \end{itemize}
\end{abstract}

\section{Introduction}

Proof assistants have become invaluable tools in the hands of mathematicians, computer scientists and 
proof engineers that aim to build certified theories, software, systems and hardware, as evidenced by
successful, large formalization projects ranging from famously hard mathematical results
(\cite{DBLP:conf/itp/GonthierAABCGRMOBPRSTT13},\cite{DBLP:journals/corr/HalesABDHHKMMNNNOPRSTTTUVZ15})
to realistic compilers (\cite{Leroy-compcert-06}, \cite{DBLP:journals/jfp/TanMKFON19}), 
program logics (\cite{DBLP:journals/cacm/JungJKD21,DBLP:books/daglib/0034962}), operating systems
(\cite{DBLP:conf/sosp/KleinEHACDEEKNSTW09,DBLP:conf/osdi/GuSCWKSC16}) and even hardware design
(\cite{DBLP:journals/cj/AkbarpourATH10,10.1145/3314221.3314622,DBLP:conf/cpp/Chlipala20}). Ultimately, 
however, all these 
formalizations rely on a Trusted Theory Base (TTB), that consists of the mathematical foundations of 
the proof-assistant -most often a variant of Higher-Order Logic, Set Theory or Type Theory- 
and a Trusted Code Base (TCB): its actual implementation in a general purpose programming language. 
To obtain the highest guarantees on the proof assistants results, one should in principle also verify the consistency of the foundations,
usually by building models in a Trusted Theory (Zermelo-Fraenkel Set Theory being the most common one), 
the adequacy of the implementation with this theory, and the correct compilation of this implementation.

\subsection{A little history}

To date, only the HOL family of provers have benefitted from such a guarantee, due to the seminal work of 
Kummar \textit{et al} \cite{DBLP:journals/jar/KumarAMO16}, 
who built a self-formalization of Higher-Order Logic modeled by set theory (ZF) in 
HOL Light (building on Harrison's work \cite{harrison-holhol}), implemented itself in CakeML.
In contrast, for dependent type theories at the basis of proof assistants like Coq, Agda, Idris or Lean, 
self-certification, or more informally type-theory eating itself \cite{DBLP:journals/entcs/Chapman09} is 
a long-standing open problem. A rather large fragment of the theory at the basis of Coq, the Calculus of
Constructions with universes, was formalized in Coq by Barras during his PhD \cite{Barras99} and extended 
in his habilitation thesis \cite{barras-habilitation}, culminating in a proof of Strong Normalization (SN) 
and hence relative consistency with IZF with a number of inaccessible cardinals for this theory. 
His development includes a proof of subject reduction and a model-theoretic proof using a 
specification of Intuitionistic Zermelo Fraenkel Set Theory in Type Theory, 
resulting from a long line of work relating the two \cite{alti:phd93,wernerTACS97}.
Due to Gödel's second incompleteness theorem, one can only hope to prove the consistency of 
the theory with $n$ universes in a theory with $n+1$ universes. These results prove that a 
quite large fragment of Coq can be shown relatively consistent to a standard foundation.
Since Barras' work, both pen-and-paper and formalized model-theoretic proofs have been 
constructed for many variants of dependent type theories, from decidability of type-checking 
for a type theory with universes \cite{DBLP-journals/pacmpl/0001OV18} or canonicity \cite{Huber:2019uf} 
and normalization \cite{DBLP:journals/corr/abs-2101-11479} for cubical type theories.
We hence consider Coq's core type theory to be well-studied, a most recent reference for a consistency 
proof of the calculus with inductive types, universe polymorphism and cumulativity is 
\cite{DBLP.conf/rta/TimanyS18,timany.hal-01615123}.

\subsection{Goals of the project}

The theory behind Coq's calculus, the Polymorphic Cumulative Calculus of (Co-)Inductive Constructions (\PCUIC),
is rather well-studied and well-tested now: most important fragments have accepted consistency proofs and no inconsistency 
was found in the rules that have been implemented in the last 30 years. That is, until a new axiom like Univalence breaks implicit
assumptions in the calculus as happened recently for the guard checking algorithm. More worryingly,
blatant inconsistencies (accepted proofs of \coqdocind{False}), are reported at least once a year, due to 
bugs in the implementation\footnote{\url{https://github.com/coq/coq/blob/master/dev/doc/critical-bugs}}.
The source of these bugs falls generally under the category of programming errors and unforeseen interactions between 
subsystems. Coq is indeed a complex piece of software resulting from 37 years of research and development, incorporating
many features in a single system. Its architecture was last revised in the V7 series  (1999-2004) by Jean-Christophe 
Filliâtre \cite{filliatr2020coqws,filliatre:hal-02890416}, following the de Bruijn criterion. It means that \Coq does have a well-delimited, trustable proof-checking kernel, 
so these errors are not to be attributed to bad design in general. Rather, the problem is that this "small" kernel 
already comprises around 20kLoC of complex \OCaml code + around 10kLoC of C code implementing a virtual machine for conversion.
The system also relies on the whole OCaml compiler when used with the \texttt{native_compute} tactic for fast computation/conversion.
To be fair (and grateful!), one should note that we never had to blame OCaml for a miscompilation resulting in an inconsistency.
In conclusion, to avoid these errors, we should rather apply program verification to \Coq's implementation.

This is a slightly different endeavor than the above works. Indeed, mechanized or not, the aforementionned proofs are 
concerned with idealized versions of the calculus that do not correspond to the actual implementation of PCUIC in OCaml,
nor do they have any bearing on its compiled version, \textit{a priori}. The \MetaCoq\footnote{\url{https://metacoq.github.io}} project's
goal is to bridge the gap between the model theoretic justification of the theory and 
the actual implementation of the \Coq kernel. To do so, we need to answer the following questions in a formal, concrete way:

\begin{itemize}
\item What calculus is implemented by Coq exactly?
\item Which meta-theoretical properties hold on the implementation?
\end{itemize}

To answer these questions, we develop a \emph{specification} of \Coq's type theory (\S \ref{sec:syntax-sem}), 
a \Coq definition of a type-checker and conversion procedure that corresponds to the current \Coq implementation, and verify both 
the sanity of the specification and correctness and completeness of the implementation.

\paragraph{Plan of the article.}
To verify the sanity of the specification we develop the meta-theory (\S \ref{sec:metatheory}) of the PCUIC calculus
and show that it enjoys type preservation (\S\ref{sec:SR}) and principality (\S\ref{sec:principality}) of types, along with 
the expected confluence of reduction (\S\ref{sec:confluence}) and proof that conversion is a congruence.
We can then construct a corresponding type checker (\S \ref{sec:typechecker})
that is shown correct and complete with respect to the specification. Finally, to be able 
to execute this type-checker on large examples, we can extract it to OCaml and compare it to the type-checker 
implemented in \Coq. Ideally, this last step should also be verified: we present a verified erasure procedure (\S \ref{sec:erasure})
that takes a \Coq environment and definition and produces a program in an (untyped) weak call-by-value 
$\lambda$-calculus extended with a dummy $\square$ constructor ($\lambda_\square$). We prove that erasure preserves
observations of values, starting from a well-typed closed term. In combination with the \CertiCoq compiler \cite{certicoq.CoqPL} 
from $\lambda_\square$ to \texttt{C-light} and the \textsc{CompCert} \cite{compcert} compiler from 
\texttt{C-light} to assembly, this will provide an end-to-end verified implementation of \Coq's kernel.
We discuss the remaining hurdles to realize this and further extensions in section \ref{sec:future}.

\paragraph{Attribution} 
The results surveyed in this article are due to \href{https://metacoq.github.io/#team--credits}{\MetaCoq team}
as a whole: Abhishek Anand, Dannil Annenkov, Simon Boulier, Cyril Cohen, Yannick Forster, Meven Lennon-Bertrand, 
Gregory Malecha, Jakob Botsch Nielsen, Matthieu Sozeau, Nicolas Tabareau and Théo Winterhalter.

\paragraph{Link to the formal development}
This article is best read online, as links in the text point to the formal development definition, which are
generally too large to include in the presentation.

\section{Syntax and Semantics}
\label{sec:syntax-sem}

\def\constr#1{\coqdocconstructor{#1}}

\begin{figure}
  \begin{center}
  \includegraphics[width=300px]{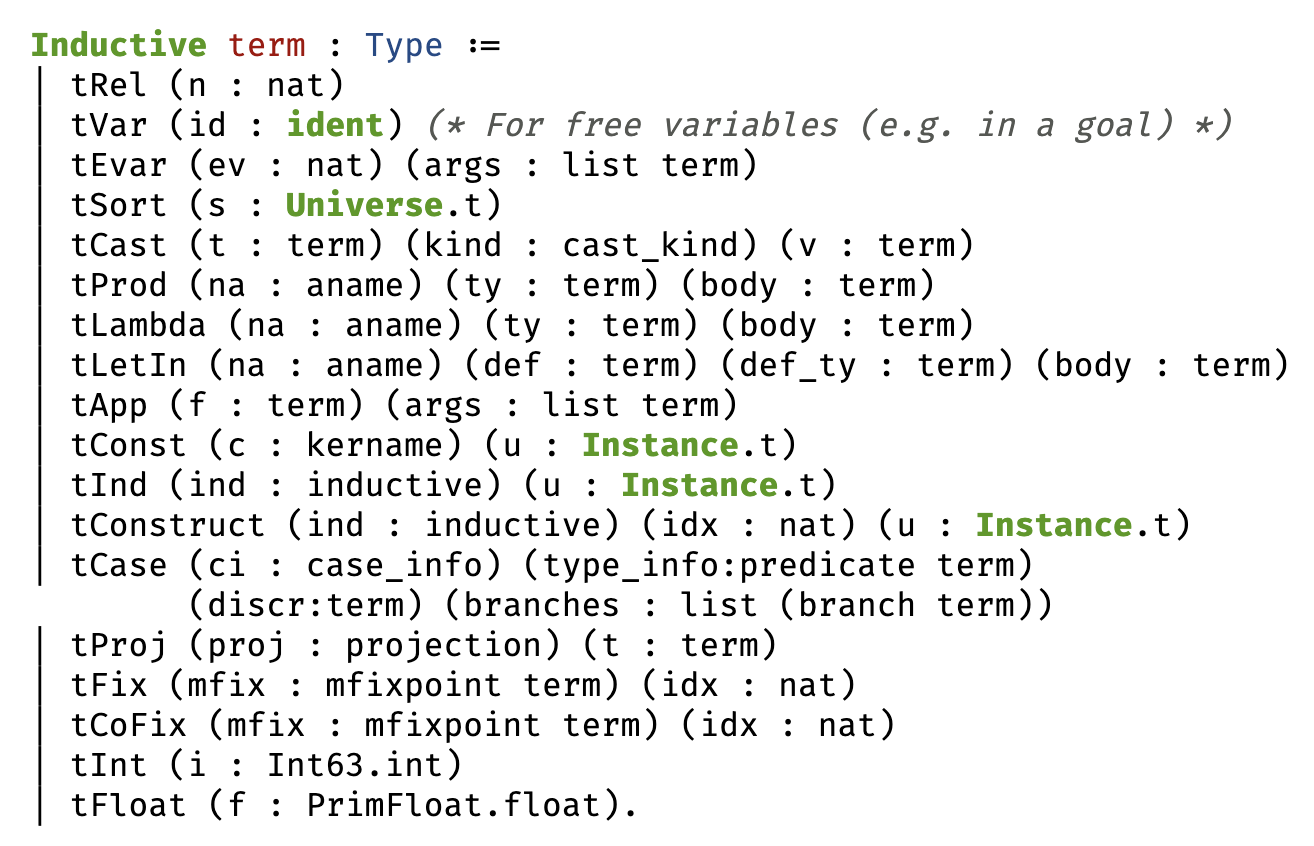}
  \end{center}
  \caption{Term syntax}
  \label{fig:term}
\end{figure}

The \MetaCoq project initially started as an extension of Gregory Malecha's \-\TemplateCoq plugin, 
developed during his PhD on efficient and extensible reflection in dependent type theory 
\cite{malecha2015thesis}. \TemplateCoq provided a \emph{reification} of \Coq's term and environment 
structures in \Coq itself, i.e. \Coq~\coqdockw{Inductive} datatype declarations that model the 
syntax of terms and global declarations of definitions and inductive types, as close as possible to 
the \OCaml definitions used in \Coq's kernel. In addition, it implemented in an \OCaml \Coq plugin 
the meta-definitions of quoting and unquoting of terms between the two representations, similarly 
to \Agda's reflection mechanism. The syntax of \href{https://github.com/MetaCoq/metacoq/blob/bc6c2a942f9dc09b7602777ba20113a7c5090982/template-coq/theories/Ast.v#L37}{terms} is given in figure \ref{fig:term}. It corresponds
closely to \Coq's internal syntax, handling local variables (\constr{tRel}, with a de Bruijn index),
free named variables (\constr{tVar}), existential variables (\constr{tEvar}), sorts (\constr{tSort}),
the type-cast construct (\constr{tCast}), dependent products, lambda-abstraction, let-ins and n-ary
application (\constr{tApp}), application of global references: constants, inductives, and constructors,
a dependent case analysis construct (\constr{tCase}, see \ref{sec:case-repr} for more details) and 
primitive projections (\constr{tProj}), fixpoints and co-fixpoints and finally, primitive integers and 
floating-point values (a recent \cite{DBLP:conf/itp/BertholonMR19} addition to \Coq).

On top of the term syntax, the \href{https://github.com/MetaCoq/metacoq/tree/bc6c2a942f9dc09b7602777ba20113a7c5090982/template-coq/theories}{\coqdocmodule{MetaCoq.Template}}
library also defines the type of local contexts which are lists of typing assumptions or local definitions, and global contexts:
associative lists of global reference names to constant/axiom or inductive and constructors declarations.

\subsection{The Template Monad}
\label{sec:template-monad}

On top of these syntactic objects, one can define an API much like \Coq's OCaml API to interact with 
the kernel of \Coq: adding and looking up global definitions, calling the type-checker or higher-level 
primitives like proof-search for specific type-class instances. In \cite{metacoqproject}, we show how
this can be organized by defining a general \coqdoccst{TemplateMonad} type-class that describes the structure 
of programs interacting with \Coq's kernel. We need a monadic abstraction to encapsulate the handling of the 
global state of the \Coq system that is being manipulated: updates to the global environment but also the 
obligation handling machinery of \Program \cite{sozeau.Coq/Russell/article}.
An interpreter for actions in this monad (e.g. adding a new definition with a given name and proof term)
is meta-programmed in \OCaml, using continuations to handle the interactive aspect of execution and 
maintaining coherence of the system state, similarly to the proof engine monad of the \textsc{MTac2}
\cite{journals/pacmpl/KaiserZKRD18} tactic language.

Using this monad, one can meta-program plugins in \Coq that take input from the user, ask for user to 
provide definitions or proofs of particular types and update the environment from computed definitions.
This can be used to define user-extensible translations \cite[\S 4]{metacoqproject}, or to derive 
lenses for a record definition \cite[\S 5]{metacoqproject}. The \coqdoccst{TemplateMonad} has a variant
that also allows \emph{extraction} of the monadic programs so they can be run efficiently in \OCaml, putting
it on par with the development of plugins for \Coq directly in \OCaml. \MetaCoq plugins deriving 
parametricity theorems \cite[\S 4.3.1]{metacoqproject} and induction principles and subterm relations from inductive 
definitions \cite{liesnikov2020} can be defined this way, opening the possibility to verify their implementations.
For parametricity for example, one can show that there is a procedure to construct from any well-typed term 
from \Coq a proof that the parametricity predicate derived from its type holds on the term.
As shown by Pédrot and Tabareau in \cite{DBLP:journals/pacmpl/PedrotT20}, this can in turn be used 
to build internal syntactic models of richer, effectful type theories. 

In the rest of this article, we will review how we built certified foundations needed for such efforts,
that is the typing specification and the type inference algorithm used to check well-typedness.

\subsection{Typing, Reduction and Conversion}
\label{sec:typing}

The calculus at the basis of \Coq is the Polymorphic Cumulative Calculus of (Co-)Inductive Constructions (\PCUIC).
\PCUIC is a general dependently-typed programming language, with pattern-matching and (co-)recursive definitions,
universe polymorphic global declarations (constants or inductive types). Its origin is the Calculus of 
Constructions of Coquand and Huet \cite{COQUAND88} with $\beta\eta$-conversion, 
extended by the (Co-)\-Inductive type definition scheme of Paulin-Mohring \cite{paulinTLCA93}, guarded (co-)fixpoint 
definitions \cite{DBLP-conf/icalp/Gimenez98}, universe polymorphism \cite{DBLP-conf/itp/SozeauT14}
and cumulative inductive types \cite{DBLP.conf/rta/TimanyS18}. The latest additions to the core calculus are a definitionally 
proof-irrelevant sort \coqdockw{SProp} and the addition of primitive types \cite{armand2010itp}.
While they are supported in the syntax, they are not yet supported by our specification.

The sort system includes an infinite hierarchy of predicative universes \Type{i} $(i ∈ \mathbb{N})$ and 
an impredicative sort \Prop. We consider \coqdockw{Set} to be a synonym for \Type{0}, hence its interpretation 
is always predicative\footnote{\Coq supports an \texttt{-impredicative-set} flag to switch to an impredicative interpretation,
but it is seldom used today}. A specificity of \Coq is the treatment of \Prop and in particular the singleton elimination
rule that allows to eliminate propositional content into computational content, if it is trivial (a proof of absurdity,
an equality, conjunction or accessibility proof): we will see in the section on erasure (\ref{sec:erasure}) how that 
is justified from the computational standpoint.

\subsubsection{Conversion, Cumulativity}

In dependent type theory, conversion and typing are usually intertwined, through the conversion rule 
which states that a term of type $T$ can be seen as a term of type $U$ for any (well-formed) typed $U$
convertible to $T$. \PCUIC is presented in the style of Pure Type Systems, where this conversion relation 
is untyped and can be defined independently on raw terms as the reflexive, symmetric and transitive closure of one-step reduction. 
We hence first define a \href{https://github.com/MetaCoq/metacoq/blob/bc6c2a942f9dc09b7602777ba20113a7c5090982/template-coq/theories/Typing.v#L145}{reduction relation}
as an inductive predicate that includes all reduction rules of \PCUIC: $\beta$ for application, ι for cases, 
$\zeta$ for let-ins, \texttt{fix} and \texttt{cofix}, \texttt{delta} for constants and congruence rules allowing
to apply reductions under any context (under lambdas, dependent products, etc). We take its closure with an additional twist:
rather than a strict reflexivity rule, we define an $\alpha$-equivalence relation that ignores the name annotations of
binders (they are included in the core syntax for printing purposes only). Moreover, this relation is parameterized by 
a relation on universes to implement syntactic cumulativity of universes and inductive types. 
Indeed in \Coq's theory we have: \[ \frac{i ≤ j}{\Type{i} \leq \Type{j}}\]

A similar rule applies to cumulative inductive types. We call this relation $\alpha$-cumulativity when instantiated
with the large inequality of universes, in which case it is only a preorder.
In case we instantiate it with equality of universes, we recover the usual $\alpha$-conversion relation, 
which is an equivalence.

Two terms are hence in the cumulativity relation if they can be linked by reductions \emph{or expansions} up-to 
$\alpha$-cumulativity.

\subsubsection{Typing}

The \href{https://github.com/MetaCoq/metacoq/blob/f4e9a80ea336fc7154aac9cce5385bd36a48125a/pcuic/theories/PCUICTyping.v#L332}{typing relation} 
of \PCUIC is a fairly standard inductively defined relation that mostly corresponds to usual 
"on paper" treatments (e.g. \Coq's reference manual \cite{coq813}):

\begin{figure}[h]
\includegraphics{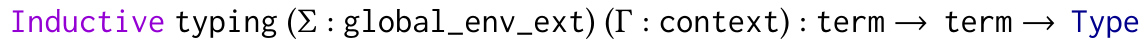}

\includegraphics{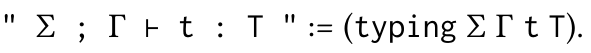}
\caption{Type signature and notation for typing}
\end{figure}

The typing judgement is a predicate taking as parameters the global context extended with a universe declaration, 
a local context of assumptions and two \coqdocind{term}s corresponding to the subject and type of the typing rules.
Derivations are defined in the \Typea sort to allow for easy inductions on the size of derivations. The typing
rules are explained in detail in \cite[\S 2]{metacoqproject} and \cite{coqcoqcorrect}[\S 2.2]. The typing rules
are syntax-directed, i.e. there is one rule per head constructor of \texttt{term}, \emph{except} for the 
cumulativity rule which can apply anywhere in a derivation. Note that we use a standard de Bruijn encoding for local variables, along with lifting and parallel substitution 
operations. As can be seen from the definition of figure \ref{fig:term}, we used a nested datatype definition 
(\texttt{list term}, \texttt{list (branch term)}), hence some care must be taken to define recursive definitions 
and user-friendly induction principles on \href{https://github.com/MetaCoq/metacoq/blob/master/template-coq/theories/Induction.v}{terms}
and \href{https://github.com/MetaCoq/metacoq/blob/bc6c2a942f9dc09b7602777ba20113a7c5090982/template-coq/theories/Typing.v#L1057}{derivations},
lifting a predicate on terms to lists in the appropriate way. This is done by defining first a size measure on terms 
and then using well-founded induction on sizes of terms and derivations to derive easy to use induction principles.

\paragraph{Global environments} 
Typing extends straightforwardly to local and global contexts. We formalize in particular which invariants should hold
on the definition of inductive types, including strict positivity and the invariants enjoyed by cumulative inductive types.
This is one point where we depart from the pen-and-paper treatments: indeed in \cite{DBLP.conf/rta/TimanyS18}, the theory that is studied is
an idealization of \Coq's implementation where equality is a judgement and inductive declarations do not carry parameters. 
In contrast, the implementation cannot rely on typing conditions to decide the cumulativity judgement and subtyping is rather 
defined on two different instances of the \emph{same} inductive type, e..g \texttt{list@\{Set\} nat} and \texttt{list@\{i\} nat}.
We hence had to retro-engineer, from \Coq's \OCaml code, a proper treatment of cumulative inductive types. We'll see in section 
\ref{sec:case-repr} that this was a non-trivial endeavor.

\subsection{Translation from Template to PCUIC}
\label{sec:translations}

The \constr{tApp} constructor represents n-ary application of a term $f$ to a list of arguments $args$. 
This follows rather closely \Coq's implementation, where the application node takes an array of arguments,
for an even more compact representation of applications. Immediate access to the head of applications is 
an important optimization in practice, but this representation, while memory-efficient, imposes a hidden
invariant on the term representation: the term $f$ should not itself be an application, and the list of 
arguments should also always be non-empty. The application typing rule is likewise complicated as we have
to consider application to a spine of arguments, rather than a single argument.

In \Coq's kernel, this is handled by making the \coqdocind{term} type abstract and using smart constructors 
to enforce these invariants. Replicating this in \Coq is tedious, as we have to either:
\begin{itemize}
\item work everywhere with an abstract/subset type of terms, precluding the use of primitive fixpoint
  and case-analysis
\item or work with the raw syntax and add well-formedness preconditions everywhere
\end{itemize}

Our solution is to interface with \Coq using the raw \coqdoccst{Template} \coqdocind{term} syntax, 
keeping close to the implementation. To avoid dealing with well-formedness assumptions, we define a 
translation from this syntax to the \PCUIC~\coqdocind{term} syntax where application is a binary constructor.
We define similar reduction and typing judgments on the simplified syntax and show the equivalence of the 
two systems \coqdoccst{Template} and \PCUIC. This crucially relies on well-founded induction on the size 
of derivations to "reassociate" between binary applications and n-ary ones.
The metatheory hereafter is developed on the more proof-friendly \PCUIC syntax, but its theorems also apply to the original system. 
We simplify the \PCUIC syntax further by 
removing the \constr{tCast} constructor and translating it by an application of the identity function: 
this is observationally equivalent. The cast syntax in \Coq is solely used to mark a place where a specific 
conversion algorithm should be used to convert the inferred type of a term with a specified expected type.
This is used to call \texttt{vm\_compoute} or \texttt{native\_compute}
to perform the cumulativity check, rather than \Coq's standard "call-by-need" conversion algorithm.
As we do not formalize these fast conversion checks, this is not a loss. Note also
that this has no bearing on the typeability of the term, \emph{in theory}. Only \emph{in practice} performing 
conversion with the default algorithm might be infeasible.

\section{Metatheory}
\label{sec:metatheory}

Now armed with the definition of typing and reduction in \PCUIC, we can start proving the usual metatheoretical results 
of dependent type theories. We first derive the theory of our binding structures: the usual lemmas about de Bruijn operations
of lifting and substitution are proven easily. 

\subsection{Digression on binding}

Unfortunately, at the time we started the project (circa 2017), 
the Autosubst framework \cite{DBLP-conf/itp/SchaferTS15} could not be used to automatically derive this theory for us, 
due to the use of nested lists. We however learned from their work \cite{DBLP-conf/cpp/StarkSK19} and developed the more expressive σ-calculus,
defined from operations of renaming (for a renaming $\mathbb{N} → \mathbb{N}$) and instantiation (for a function $\mathbb{N} → \coqdocind{term}$),
which provide a more proof-friendly interface to reason on the de Bruijn representation. We show that \Coq's kernel 
functions of lifting and substitution (carrying just the number of crossed binders) are simulated with 
specific renaming and instantiation operations. Both variants are still of interest: it would be
out of the question to use the σ-calculus operations which build closures while traversing terms 
in \Coq's implementation. However the structured nature of the $\sigma$-calculus and its 
amenability to automation, having a decidable equational theory \cite{DBLP-conf/cpp/StarkSK19}, 
is a clear advantage in practice.

One example where this shines is the treatment of (dependent) let-bindings in the calculus. Dependent let-bindings 
are an entirely different beast than ML-like let-bindings which can be simulated with abstraction and application.
In particular, three rules of reduction can apply on local definitions:
\def\kw#1{\coqdockw{#1}\xspace}
\[\begin{array}{lclll}
    Γ \vdash \kw{let}~x := t~\kw{in}~b & \rightsquigarrow & b[t/x] & & \zeta \\ 
    Γ \vdash \kw{let}~x := t~\kw{in}~b & \rightsquigarrow & \kw{let}~x := t~\kw{in}~b' & 
    \texttt{ when~} Γ , x := t \vdash b \rightsquigarrow b' & \texttt{cong-let-body} \\
    Γ, x := t, Δ \vdash x & \rightsquigarrow & \uparrow^{|Δ|+1}(t) & & δ
\end{array}\]

Here $\uparrow^{n}(t)$ represents the shifting of indices of the free variables of $t$ by $n$, and 
$b[t/x]$ the usual capture-avoiding substitution. The first rule is usual let-reduction, the second is 
a congruence rule allowing to reduce under a \coqdockw{let} and the last allows to expand a local definition
from the context. In the course of the metatheoretical development
we must prove lemmas that allow to "squeeze" or smash the let-bindings in a context. This results in a 
reduced context with no let-bindings anymore and a corresponding substitution that mixes the properly substituted
bodies corresponding to the removed let-bindings and regular variables, to apply to terms in the original
context. This involves interchanging $δ$, $zeta$ and $\texttt{cong-let-body}$ rules combined with 
the proper liftings and substitutions. This kind of reasoning appears in particular as soon as we want to 
invert an application judgment as let-bindings can appear anywhere in the type of the functional argument. 
Using $σ$-calculus reasoning and building the right abstractions tremendously helped simplify the proofs 
which would otherwise easily become indecipherable algebraic rewritings with the low level indices in 
liftings and substitutions.

\subsection{Properties}
\label{sec:properties}

\subsubsection{Structural Properties}
\label{sec:structural-properties}

The usual \href{https://github.com/MetaCoq/metacoq/blob/bc6c2a942f9dc09b7602777ba20113a7c5090982/pcuic/theories/PCUICWeakening.v#L1095}{Weakening}
and \href{https://github.com/MetaCoq/metacoq/blob/bc6c2a942f9dc09b7602777ba20113a7c5090982/pcuic/theories/PCUICSubstitution.v#L2696}{Substitution} 
theorems can be proven straightforwardly by induction on the derivations. We use a beefed-up eliminator that automatically lifts the typing 
property we want to prove to well-formedness of the global environment, which also contains typing derivations. Likewise, we always
prove properties simulataneously on well-formed local contexts and type derivations, so our theorems provide a conjunction of properties.
When we moved to the σ-calculus represenation, we factorized these proofs by first stating renaming and instantiation lemmas,
from which weakening and substitution follow as corrolaries. We also verify that typing is  
\href{https://github.com/MetaCoq/metacoq/blob/bc6c2a942f9dc09b7602777ba20113a7c5090982/pcuic/theories/PCUICAlpha.v#L261}{invariant by alpha-conversion},
so name annotations on binders are indeed irrelevant.

\subsubsection{Type Preservation}
\label{sec:SR}

Proving subject reduction (a.k.a. type preservation) for dependent type theories can be rather difficult in a setting 
where definitional equality is typed, as it usually requires a logical relation argument/model construction, see e.g.
\cite{DBLP-journals/pacmpl/0001OV18}. However, the syntactic theory is relatively well-understood for PTS: one can 
first independently prove context conversion/cumulativity and injectivity of $\Pi$-types (i.e. $Π x : A. B ≡ Π x : A'. B' \rightarrow A ≡ A' \wedge B ≡ B'$),
to prove type preservation in the application case. Similarly, we have injectivity of inductive type applications, 
up-to the cumulativity relation on universes. 

However, two other difficulties arise for \PCUIC. First, we are considering a realistic type theory, will full-blown 
inductive family declarations, with a distinction between parameters and indices (that can contain let-bindings), 
and cumulative, universe polymorphic inductive types. To our knowledge, nobody ever attempted to formalize the 
proof at this level of detail before, even on paper. There is a good reason for that: the level of complexity 
is very high. Showing that the dependent case analysis reduction rule is sound mixes features such as let-bindings, 
the de Bruijn representation of binding adding various liftings and substitutions and the usual indigestible 
nesting of quantifications on indices in the typing rules of inductive types, constructors and branches. 
This is also ample reason to verify the code: many critical bugs a propos let-binding and inductive types
were reported over the years. To tame this complexity, we tried to modularize the proof and provide the most 
abstract inversion lemmas on typing of inductive values and \kw{match} constructs. 

Second, \Coq's theory is known to be broken regarding positive co-inductive types and subject reduction.
We hence parameterize the proof: subject reduction holds only on judgments where no dependent case analysis
is performed on co-inductive types. Negative co-inductive types implemented with primitive projections however
can be show to enjoy subject reduction without restriction.

\subsubsection{Confluence}
\label{sec:confluence}

To support inversion lemmas such as $\Pi$-type injectivity, we need to show that reduction is confluent.
From this proof, it follows that the abstract, undirected conversion relation $T ≡ U$ is equivalent to 
reduction of the two sides to terms $T'$ and $U'$ that are in the syntactic α-cumulativity
relation. We extend here Takahashi's refinement \cite{journals/iandc/Takahashi95} of Tait/Martin-Löf's 
seminal proof of confluence for $\lambda$-calculus. The basic idea of the confluence proof is to consider
parallel reduction instead of one-step reduction and prove the following triangle lemma:

\begin{center}
  \includegraphics{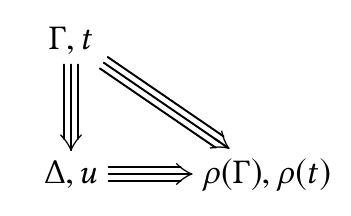}
\end{center}

The $\rho$ function here is an "optimal" reduction function that reduces simultaneously all 
parallel redexes in a term or context. The fact that we need to consider contexts is due 
to let-bindings again: in one step of $\rho$, we might reduce a local definition
to an abstraction, expand it in function position of an application and reduce the produced beta-redex.
Using the triangle property twice, it is trivial to derive the diamond lemma and hence confluence for parallel reduction.
By inclusion of one step reduction in parallel reduction, and parallel reduction in the transitive closure of one-step 
reduction (the "squashing" lemma), reduction is also confluent. This last part of reasoning is done abstractly,
but accounting for the generalization to reduction in pairs of a context and a term.
It then suffices to show commutation lemmas for reduction and $\alpha$-cumulativity to show the equivalence 
of reduction up-to $\alpha$-cumulativity and the undirected cumulativity relation.
Confluence is crucial to show transitivity of the directed version.

Using this characterization of cumulativity, we can easily show that it is a congruence and that it enjoys
expected inversion lemmas: if two $\Pi$-types are convertible, then they both reduce to $\Pi$-types that are
in the $\alpha$-cumulativity relation, so their domains are convertible and their codomains are in the cumulativity 
relation. Similarly, $\Pi$-types cannot be convertible to sorts or inductive types: there is no confusion between
distinct type constructors.

\subsubsection{Principality}
\label{sec:principality}

As \PCUIC is based on the notion of cumulativity rather than mere conversion, type uniqueness does not hold 
in the system. However, the refined property of principality does: for any context $Γ$ and well-typed term $t$,
there is a unique type $T$ such that any other typing of $t$, $Γ ⊢ t : U$ we have $T ≤ U$. 
This property is essential for deciding type-checking: to check $Γ ⊢ t : U$, it suffices to infer the principal type of $t$ 
and check cumulativity. Principal typing is also used by the erasure procedure to take sound decisions based 
on the principal type of a given term.

\subsection{Strengthening}
\label{sec:strengthening}

\def\FV#1{\texttt{FV}(#1)}
The last expected structural property of the system is strengthening (a.k.a. thinning), 
which can be stated as:

\[Γ, x : A, Δ ⊢ t : T → x ∉ \FV{\Delta} \cup \FV{t} \cup \FV{T} → Γ, Δ ⊢ t : T\]

This property ensures that typing is not influenced by unused variables, and is at the basis of 
the \texttt{clear} tactic of \Coq, a quite essential building block in proof scripts. Unfortunately,
this property \emph{cannot} be proven by a simple induction on the typing derivation: 
the free-floating conversion rule allows to go through types mentioning the variables to clear,
even if they do not appear in the term and type in the conclusion of the whole derivation.

\subsection{Bidirectional Type Checking To The Rescue}
\label{sec:bidir}

This unfortunate situation can be resolved by reifying the principality property and type checking
strategy as a bidirectional typing inductive definition. This variant of typing explicitly keeps track
of the information flow in typing rules. In practice it separates the syntax-directed rules in a
synthesis (a.k.a. inference) judgment (the term position is an input, while the type is an output)
from the non-directed ones as checking rules (both positions are input). In \cite{conf/itp/Lennon-Bertrand21}, 
Meven Lennon-Bertrand develops a bidirectional variant for \PCUIC, show equivalent to the orignal \PCUIC,
in which strengthening and principality become trivial to derive. 

The crux of these argument is that bidrectional typing derivations are "canonical" for a given term,
and move uses of the conversion rule to the top of the derivation, where they are strictly necessary.
In addition, multiple cumulativity rules get "compressed" into a single change-of-phase rule, relying on 
transitivity of cumulativity. In a bidirectional synthesis derivation, if a variable does not appear 
in the term position, then it cannot appear in the inferred type. 
Simultaneously, in a checking derivation, if a variable does not appear in the term and type, 
then it cannot appear in these positions in any of the subderivations.

\subsection{Case In Point}

This detour through bidirectional typechecking is not accidental. In \cite{coqcoqcorrect}, 
we only proved the soundness of a typechecking algorithm for \PCUIC (\S \ref{sec:typechecker}).
It is in the course of formalizing the completeness of the type-checker (\S \ref{sec:typechecker}) 
that we discovered a problem in the typing rules of \Coq. The problem appears in the dependent case analysis 
construct \coqdockw{match}. The gist of the typing rule was to typecheck the scrutinee
at some unspecified inductive type $\coqdocind{X}@\{ \vec{u} \}~\vec{p}~\vec{i}$, 
where $\vec{u}$ is a universe instance,$\vec{p}$ the parameters and $\vec{i}$ the 
indices of the inductive family. The \texttt{match} construct also takes an elimination
predicate, expected to be of type:
\[\Pi (\overrightarrow{x : I@[\vec{v}]}), \coqdocind{X}@\{\vec{v}\}~\vec{p'}~\vec{x} \rightarrow \Typea\]

Looking at this type, we would extract the universe instance $\vec{v}$ and 
parameters $p'$ of the inductive \coqdocind{X} assumption.
The typing rule checked that the universe instance of the scrutinee $\vec{u}$ was convertible
to $\vec{v}$, rather than only in the cumulativity relation according to the subtyping rules 
for cumulative inductive types. It also compared the parameters $\vec{p}$ and $\vec{p'}$
for convertibility, by first lifting $\vec{p}$ in a context extended with the $\overrightarrow{x : I@[\vec{v}]}$
bindings, but these two instances did not necessarily live in the same type!

These mistakes lead to a loss of subject reduction, if cumulativity is used to lower the universes of the scrutinee, 
making the whole pattern-matching untypeable\footnote{\url{https://github.com/coq/coq/issues/13495}}. 
The problem appeared immediately while trying to prove completeness of type-checking, at the stage of designing the
bidirectional typing rules: the flow of information was unclear and led us to the bug.
We also realized that, to justify the comparison of the parameters, we would need  
to verify that $\vec{x} ∉ \FV{\vec{p'}}$ and apply strengthening, which as we have just seen is not 
directly provable on undirected typing rules. This motivated us to push for a change in the term 
representation of \texttt{match} in 
\Coq\footnote{\href{https://github.com/coq/ceps/blob/master/text/inductive-branch-predicate-representation-and-reduction.md}{CEP 34} by Hugo Herbelin, \href{https://github.com/coq/coq/pull/13563}{Coq PR 13563} by Pierre-Marie Pédrot, integrated in Coq 8.14}
that solves both problems at once, by storing at the \texttt{match} node the universe instance and parameters
that define the eliminator, and doing a sound cumulativity test of the infered type of the scrutinee and the 
(reconstructed) expected type of the eliminator. We are currently finishing to update the whole \MetaCoq development
to handle this change of representation\footnote{\url{https://github.com/MetaCoq/metacoq/pull/534}}.

\label{sec:case-repr}
\section{A Type-Checker for PCUIC}
\label{sec:typechecker}

\subsection{Cumulativity}

In \cite[\S 3]{coqcoqcorrect}, we present a sound type-checker for \PCUIC. To implement type-checking, 
we had to first develop a reasonably efficient reduction machine and algorithms to decide 
cumulativity. There are three separate algorithms at play to implement the cumulativity test.

\paragraph{Universes} in \Coq are floating variables subject to constraints \cite{HerbelinUniverses}, 
not directly natural numbers. To ensure consistency, one hence needs to verify that the 
constraints always have a valuation in the natural numbers. 
This boils down to deciding the (in-)equational theory of the tropical algebra
$(\mathbb{N}, \mathtt{max}, + k, \leq)$. We develop a longest-simple-paths algorithm to check 
consistency of universe constraints: the valuation of each variable can be read off as the weight
of its longest simple path from the source ($\Type{0}$).
This is a naïve model and implementation of the state-of-the-art algorithm implemented in \Coq,
which derives from an incremental cycle detection algorithm \cite{journals/talg/BenderFGT16}
and whose formal verification is a work-in-progress \cite{gueneau-hal-02167236}.
Our specification is more expressive than \Coq's current implementation, as it is
able to handle arbitrary $ℓ + k ≤ ℓ' + k'$ constraints between universe expressions, avoiding
to distinguish so-called \emph{algebraic} universes and implement costly universe refreshing operations
when building terms. We hope to integrate this generalization back in \Coq's implementation.
Using this consistency check, it is easy to define $\alpha$-cumulativity by structural recursion on terms.

\paragraph{Reduction} We implement a weak-head reduction stack machine that can efficiently 
find the weak-head normal form of a term. To define this function, we must assume an axiom 
of strong normalization, which implies that reduction is well-founded on well-typed terms.
This is the only axiom used in the development.

\paragraph{Conversion} 
\Coq uses a smart, mostly call-by-name, conversion algorithm, that uses performance-critical 
heuristics to decide which constants to unfold and when. 
\Coq's algorithm does not naïvely reduce both terms to normal form to compare them, 
but rather incrementally reduces them to whnfs (without δ-reduction),
compare their heads and recursively calls itself. When faced with the same defined constant
on both sides, it first tries to unify their arguments before resorting to unfolding, resulting
in faster successes but also potentially costly backtracking.

The main difficulty in the development of the conversion algorithm is that its termination and 
correctness are intertwined, so it is developed as a dependently-typed program that takes 
well-typed terms as arguments (ensuring termination of recursive calls assuming SN) and returns a 
proof of their convertibility (or non-convertibility). In other words it is proven sound and  
complete by construction. The design of the termination measure also involves a delicate 
construction of a dependent lexicographic ordering on terms in a stack due to Théo Winterhalter
\cite{winte2020}.

\subsection{Type Checking}

\begin{figure}[h]
\includegraphics[width=450px]{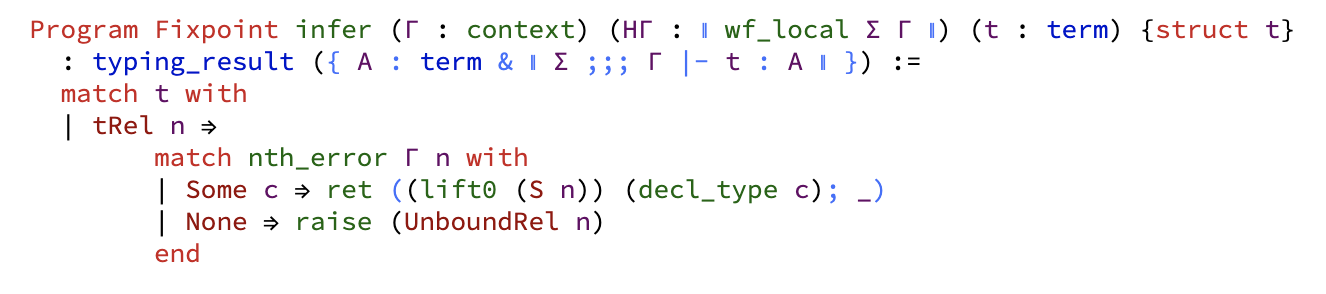}
\caption{Type inference function excerpt}
\label{fig:infer}
\end{figure}

On top of conversion, implementing a \href{https://github.com/MetaCoq/metacoq/blob/bc6c2a942f9dc09b7602777ba20113a7c5090982/safechecker/theories/PCUICTypeChecker.v#L262}{type inference algorithm} 
is straightforward:
it is a simple structurally recursive function that takes well-formed global and local contexts
and a raw term. It simply checks if the rule determined by the head of the term can apply. Figure \ref{fig:infer} 
shows the type and beginning of the inference algorithm.

Again, the function is strongly typed: its result lives in the \coqdocind{typing_result} monad,
which is an exception monad, returning (\coqdoccst{ret}) a sigma-type of an inferred type $A$ and a "squashed"
proof that the term has this type or failing with type error (\coqdoccst{raise}). As all our derivations are in $\Typea$ by 
default, we use explicit squashing into $\Prop$ when writing programs manipulating terms:

\vspace{0.6em}
\includegraphics[width=260px]{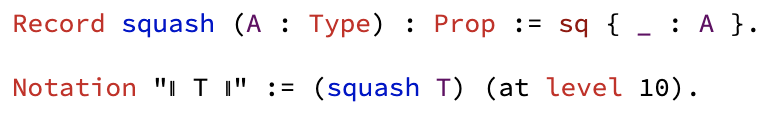}

The elimination rules for propositional inductives ensures that our programs cannot
make computational choices based on the shape of the squashed derivations, and that 
extraction will remove these arguments. The extracted version of \coqdoccst{infer} hence only 
takes a context (assumed to be well-formed) and a term and returns an error or an infered 
type, just like in \Coq's implementation.

Using the bidirectional 
presentation of the system, we can simplify the correctness and completeness proof in 
\cite{coqcoqcorrect} as the algorithm really follows the same structure as bidirectional derivations.
Type-checking is simply defined as type inference followed by a conversion test, as usual.

\subsection{Verifying Global Environments}

Once the type-checker for terms is defined, we can lift it to \href{https://github.com/MetaCoq/metacoq/blob/bc6c2a942f9dc09b7602777ba20113a7c5090982/safechecker/theories/PCUICSafeChecker.v#L2148}{verify} 
global environment declarations.
For constants and axioms, this is straightforward. However, declarations of inductive types are more
complex and require to first define a sound context cumulativity test, a strict positivity check and
to turn the universe constraints into a graph structure. This is again done using a monad \coqdoccst{EnvCheck}:

\includegraphics[width=400px]{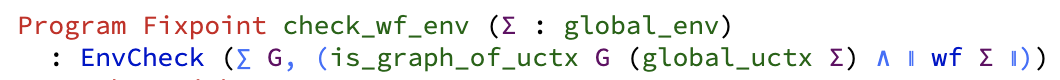}

Given a global environment $\Sigma$, this produces either an error or a pair of a graph and a proof
that the universe graph models the constraints in $\Sigma$ and a (squashed) proof that the environment is 
well-formed.

\section{Erasure from PCUIC to λ-calculus}
\label{sec:erasure}

The type-checker can be extracted to \OCaml and run on reasonably large programs. For example it can 
be used to successfully check the prelude of the HoTT library \cite{DBLP.conf/cpp/BauerGLSSS17}, including a large proof
that adjoint equivalences can be promoted to homotopy equivalences. However, our first attempt to 
extraction was unsuccessful: we had to change the \Coq definitions so that \OCaml could typecheck the
generated code, as we hit a limitation of the extraction mechanism in presence of dependently-typed 
variadic functions. The obvious next step was hence to verify the erasure procedure itself!

In \cite[\S 4]{coqcoqcorrect}, we present a sound erasure procedure from \PCUIC to untyped, 
call-by-value $\lambda$-calculus. This corresponds to the first part of \Coq's extraction mechanism
\cite{conf/types/Letouzey02}, which additionally tries to maintain simple types corresponding 
to the original program. Erausre is performed by a single traversal of the term, expected to be well-typed.
It checks if the given subterm is a type (its type is a sort $\Prop$ or $\Typea$) or if it is a 
proof of a proposition (its type $P$ has sort $\Prop$), in which case it returns a dummy $\square$ term,
and otherwise proceeds recursively, copying the structure of the original term. The result of
\href{https://metacoq.github.io/html/MetaCoq.Erasure.Extract.html}{erasure} hence contains no type information anymore, and all propositional content is replaced with $\square$.

We can prove the following correctness statement on this procedure:

\includegraphics[width=250px]{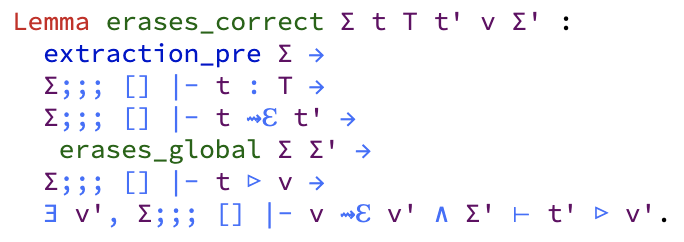}

Our correctness theorem shows that if we have a well-typed term $t$ and $t$ erases to $t'$, 
then if $t$ reduces to a \emph{value} $v$ using weak-cbv reduction, then the erased term 
$t'$ also reduces to an observationally equivalent value $v'$. The \coqdocind{extraction\_pre}
precondition enforces that the environment is well-formed. The proof follows Letouzey's original 
pen-and-paper proof closely \cite{LetouzeyPhd}. Since \cite{coqcoqcorrect}, we proved two 
additional verified passes of optimization on the erased term and environment: 
\begin{itemize}
\item We remove from $\Sigma'$ the definitions that are not used for evaluation, pruning the 
  environment from useless declarations that are no longer needed for the computation.
\item We remove dummy pattern-matchings on $\square$ terms, that should always trivially reduce 
  to their single branch.
\end{itemize}
The end result of erasure is an untyped term that contains only the raw computational content of the
original definition. It can be further compiled with the \CertiCoq compiler and \textsc{CompCert} to 
produce certified assembly code from \Coq definitions.

\section{Going further}
\label{sec:future}

We have presented the whole \MetaCoq project, which spans from the reification of \Coq terms
down to the erasure of well-typed terms to untyped $\lambda$-calculs. The whole project weights 
about ~100kLoC of \OCaml and \Coq code, and is still under active development. We think this 
is a convincing example that we can move from a Trusted Code Base consisting of \Coq's unverified 
kernel down to a Trusted Theory Base that consists of the formalized typing rules of \PCUIC and 
its axiom of Strong Normalization.

The \MetaCoq (and \CertiCoq) projects are both ongoing work subject to limitations, 
we summarize here the current state of affairs.

\subsection{Limitations}

While \PCUIC models a large part of \Coq's implementation, it still misses a few impotant features
of the theory:
\begin{itemize}
\item The $\eta$-conversion rule is not supported in our formalization, preventing us to check
  most of the standard library. Dealing with $\eta$ rules in an untyped conversion setting is 
  a notoriously hard issue. We are however hopeful that we found a solution to this problem
  by quotienting definitional equality with $\eta$-reduction, and hope to present this result 
  soon.
\item Similarly, we do not handle the new \coqdockw{SProp} sort of \Coq. Our methodology for 
  $\eta$-conversion should however also apply for this case.
\item We do not formalize yet the guard-checking of fixpoint and co-fixpoint definitions, 
  relying instead on oracles. Our strong normalization assumption hence includes an assumption 
  of correctness of the guard checkers. We are currently working on integrating a definition
  of the guard checking algorithm and verifying its basic metatheory (invariance by renaming,
  substitution, etc.).
\item We did not consider the module system of \Coq, which is mostly orthogonal to the core 
  typing algorithm but represents a significant share of \Coq's kernel implementation, we leave
  this to future work.
\item We leave out the so-called "template"-polymorphism feature of \Coq, which is a somewhat
  fragile (i.e. prone to bugs), non-modular alternative to cumulative inductive types. 
  This prevents us from checking most of the \Coq standard library today as it makes 
  intensive use of this feature. We are working with the \Coq development team to move 
  the standard library to universe polymorphism to sidestep this issue.
\end{itemize}

\subsection{Conclusion and Perspectives}

There are many directions in which we consider to extend the project:

\begin{itemize}
\item On the specification side we would like to link our typing judgment to the 
  "Coq en Coq" formalization of Barras \cite{barras-habilitation}, which provides 
  the Strong Normalization proof we are assuming, for a slightly different variant 
  of the calculus. This formalization is  based on a sized-typing discipling for 
  inductive types, which will require to show an equivalence with \PCUIC's guardness
  checker, or an update of \PCUIC to handle sized typing altogether.
\item Proving that the theory is equivalent to a variant where conversion is typed, 
  i.e. definitional equality is a judgment would also make our theory closer to 
  categorical models of type theory, e.g., Categories with Families. This can build on
  recent results in this direction by Siles and Herbelin \cite{journals/jfp/SilesH12},
  updating them to handle cumulativity.
\item In addition to the parametricity translation that we would like to prove correct,
  many syntactic models of type theory, extending it with side-effects \cite{DBLP:conf/lics/PedrotT17} 
  or forcing \cite{conf/lics/JaberLPST16} have recently been developed. \MetaCoq is the right setting
  to mechanize these results.
\item We have concentrated our verification efforts on the core type-checking algorithm of \Coq,
  but higher-level components like unification, elaboration and the proof engine would also benefit 
  from formal treatment. We hope to tackle these components in the future.
\item Finally, on the user side, we are still at the beginning of the exploration of the 
  meta-programming features of \MetaCoq. It could be used to justify for example the foundations
  of the \textsc{MTac} 2 language \cite{journals/pacmpl/KaiserZKRD18}, to turn the typed tactic language 
  into a definitional extension of \Coq's theory.
\end{itemize}

\section{Bibliography}

\bibliographystyle{eptcs}
\bibliography{lfmtp21}
\end{document}